%Paper: hep-th/9408094
%From: jedwards@husc.harvard.edu
%Date: Tue, 16 Aug 94 23:57:30 -0400

\input harvmac

\def\CA{{\cal A}}
\def\CH{{\cal H}}
\def\CO{{\cal O}}

\def\inbar{\,\vrule height1.5ex width.4pt depth0pt}
\def\IC{\relax\hbox{$\inbar\kern-.3em{\rm C}$}}
\def\IR{\relax{\rm I\kern-.18em R}}
\font\cmss=cmss10 \font\cmsss=cmss10 at 7pt
\def\IZ{\relax\ifmmode\mathchoice
{\hbox{\cmss Z\kern-.4em Z}}{\hbox{\cmss Z\kern-.4em Z}}
{\lower.9pt\hbox{\cmsss Z\kern-.4em Z}}
{\lower1.2pt\hbox{\cmsss Z\kern-.4em Z}}\else{\cmss Z\kern-.4em Z}\fi}

\def\al{\alpha}
\def\be{\beta}
\def\ga{\gamma}
\def\de{\delta}
\def\ep{\epsilon}
\def\ze{\zeta}

\def\si{\sigma}

\def\zesl{\hskip -1pt{\not{\hskip -1.5pt \zeta}}}
\def\psisl{\hskip -1pt\not{\hskip -2.5pt \psi}}
\def\psl{\hskip -2pt{\not{\hskip -2pt p}}}
\def\qsl{\hskip -1pt{\not{\hskip -1.5pt q}}\hskip -0.5pt}
\def\Rsl{{\not{\hskip -2.5pt R}\hskip 2pt}}
\def\Ysl{\hskip -1pt{\not{\hskip -1pt Y}}}

\def\hs{\hskip 2pt}

\Title{\vbox{\hbox{hep-th/9408094, HUTP--94/A022}}}
{\vbox{\centerline{Bracket Symmetries of the Classical N=1 String}}}
\vskip .2in
\centerline{Jonathan D.  Edwards}
\vskip .2in
\centerline{Lyman Laboratory of Physics, Harvard University}
\centerline{Cambridge, MA  02138, USA}
\vskip .3in
We show that it is possible to extend Moore's analysis of the classical
scattering amplitudes of the bosonic string to those of the N=1 superstring.
Using the bracket relations we are able to show that all possible amplitudes
involving both bosonic and fermionic string states at arbitrary mass levels can
be expressed in terms of amplitudes involving only massless states.
A slight generalization of Moore's original definition of the bracket also
allows us to determine the 4-point massless amplitudes themselves using only
the bracket relations and the usual assumptions of analyticity.
We suggest that this should be possible for the higher point massless
amplitudes as well.
\Date{August 1994}

\newsec{Introduction}

In a recent pair of papers
\ref\bracket{G.  Moore, ``Symmetries of the Bosonic String S-Matrix,"
YCTP-P19-93; hep-th/9310026},
\ref\addendum{G.  Moore, ``Addendum to:  Symmetries of the Bosonic String
S-Matrix," YCTP-P1-94, RU-94-19; hep-th/9404025}
Moore showed that the classical S-matrix for the flat, twenty-six dimensional
bosonic string is uniquely determined up to the string coupling constant.
Using a natural algebraic structure on the space of physical string states, he
first wrote down an infinite set of linear relations among the exact classical
scattering amplitudes for all values of the kinematical invariants.
He then used these relations to show that an arbitrary $n$-point amplitude
could be written as a linear combination of $n$-point tachyon amplitudes at
different values of the kinematical invariants.
Finally he showed that when coupled with a number of assumptions about the
analytic behavior of the scattering amplitudes, these relations could be used
to determine the $n$-point tachyon amplitude itself, thus fixing all $n$-point
amplitudes up to a choice of multiplicative constant $c_n$.
Since factorization further determines all of these potentially different
$c_n$\ in terms of the string coupling, Moore actually succeeded in uniquely
fixing the full S-matrix up to a choice of the string coupling.
The algebraic structure used by Moore was a particular case of a more general
one called the Gerstenhaber bracket, which exists for the full BRST cohomology
of an arbitrary chiral operator algebra.
This particular case is itself called the bracket, and the relations it
generates are called bracket relations.

In this paper we will show that Moore's techniques can be extended to the flat,
10-dimensional N=1 superstring.
Because of the existence of different ghost pictures, we must begin by showing
that bracket is actually well-defined.
With this done we go on to generate a set of relations among $n$-point
amplitudes involving only massless states.
Since the four possible 4-point massless amplitudes can be calculated, these
relations can be explicitly checked for $n=4$, and they indeed hold.
We next show that any $n$-point amplitude involving massive states can be
expressed in terms of one of a finite number of massless amplitudes.
Using the bracket relations generated by the massless fermionic operator--the
generalization of the supersymmetry operator to non-zero momentum--the number
of such amplitudes can be reduced to a maximal independent set.
For $n=4$\ there are two such amplitudes,
and we show that a generalization of Moore's original definition of the bracket
allows us to fix the value of both of these amplitude up to a constant.
We suggest that this should be possible for the independent higher point
massless amplitudes as well, and thus that the full S-matrix for the N=1 string
should be uniquely determined up to a choice of coupling constant.
Finally we show that the construction used by Moore in \addendum\ to lift a
restriction on $n$\ encountered in \bracket\ can also be carried over to the
$N=1$\ string.

\newsec{Review of Moore's formalism}

We begin with a review of the formalism introduced by Moore \bracket.
For the purpose of this review we will restrict ourselves to the
open bosonic string.
The extension to the closed string can be found in \bracket.
It is known that the BRST cohomology of a general chiral operator algebra
admits an operation called the Gerstenhaber bracket
\ref\gerst{B.  Lian and G.  Zuckerman, ``New Perspectives on the BRST Algebraic
Structure of String Theory," (hep-th/9211072) CMP
{\bf 154}(1993)613.}
that maps $\{*,*\}:H^{g_1}\times H^{g_2}\rightarrow
H^{g_1+g_2-1}$.
We are interested in this bracket for the case of $g_1=g_2=1$, since $H^1$\ is
just the space of physical string states.
We have the following explicit contour integral representation of the bracket:
\eqn\bdef{
\{\CO_1,\CO_2 \}(z)=\oint dw\hs (b_{-1}\CO_1)(w)
\CO_2 (z).}
Note that up to the factor of $c$\ implicitly contained in $\CO_2$
\foot{Since the physical operators are elements of the level one BRST
cohomology, each operator is actually an equivalence class of operators.
For the theory at hand we can always choose a representative of the form $cV$,
where $V$\ is a dimension one operator.},
this is nothing more than the commutator of two dimension-one operators.
We also note that since all of the operators for the flat $N=1$ string contain
a factor of the form $e^{ip\cdot X}$, and $e^{iq\cdot X(w)}
e^{ip\cdot X(z)}\sim(w-z)^{q\cdot p}e^{i(q+p)\cdot X(z)}+\cdots$, this bracket
is only defined for operators whose momenta satisfy
$q\cdot p\in\IZ$.

The bracket allows us to find relations among scattering amplitudes as follows.
We begin by choosing $n+1$\ physical state operators $V_i, i=1,\ldots,n$, and
$J$, with momenta $p_i$ and $q$, respectively.
We assume that the momenta satisfy $q+\sum_i p_i=0$\ and $q\cdot p_i\in Z$, so
that $J$\ is mutually local with respect to each of the $V_i$.
A different notation is used for the last operator since it will be used to
generate the relations among (derivatives of) the other operators.
Now consider the correlation function $\bigl<0\big|V_1(z_1)V_2(z_2)
\ldots V_n(z_n)\oint J\big|0\bigr>$, where the contour is taken around a small
circle that does not enclose any other operators.
Since we are dealing with the open string, this contour should really be
restricted to the upper-half plane, but for the moment let us imagine that $J$\
can be analytically continued so that the correlator is well-defined for a
general contour.
Since the contour does not enclose another operator, the correlator vanishes.
However if we deform the contour back around infinity, we pick up a
contribution from each of the operators equal to just the bracket of that
operator with $J$.
Finally if we fix the positions of the first, second and $n$-th operators at
the points $\infty$,1 and 0, respectively; and integrate a fixed ordering of
the positions of the remaining $n-3$\ operators over the interval $[0,1]$, we
find
\eqn\brel{
\sum_{i=1}^n(-1)^{q\cdot(p_i+\cdots+p_n)}
\CA(V_1,\cdots,\{J,V_i\},\cdots,V_n)=0,}
where $\{J,V_i\}$\ is just the bracket we defined above.
These are Moore's bracket relations or finite difference relations, so-called
because the amplitudes are evalutated at different values of the kinematic
invariants.

We must stop here to note that the above procedure can only be carried out for
scattering amplitudes involving at most twenty-six $V_i$.
This restriction is easily seen as follows.
Our ability to write down the bracket relations rests entirely upon our ability
to choose momenta such that their sum is zero, all of the operators are
on-shell, and $J$\ is mutually local with respect to each of the $V_i$.
If $d$\ is the number of non-compact target space dimensions, then for $n\le
d$\ the number of independent conditions on the momenta is equal to the number
of independent kinematical invariants available to us.
Thus we can always find the necessary momenta.
However, because of the linear relations among $d+1$\ vectors in a
$d$-dimensional space, when $n>d$\ the number of conditions is larger than the
number of invariants by $n-d$.
Thus we are no longer assured of being able to choose the necessary momenta.
In his second paper \addendum\  Moore was able to overcome this restriction by
embedding the ordinary twenty-six dimensional string in a string theory with
$26+2m$\ target space dimensions.
We shall see later that this same trick works for the $N=1$ string.

\newsec{The N=1 String}

\subsec{The Spectrum}

We begin by recalling the operator content of the $N=1$\ critical superstring
\ref\GSW{M.  Green, J.  Scwarz and E.  Witten, {\it Superstring Theory},
Cambridge University Press (1987).}.
The space $\CH$\ of physical operators is a direct sum of two subspaces,
$\CH=\CH_b\oplus\CH_f$, whose elements have either bosonic or fermionic target
space statistics, respectively.
Each of these subspaces carries both a discrete and a continuous grading, and
may be written as
\eqn\spec{
\CH_*=\bigoplus_{n\in Z_+}\int_{p\in R^{1,9}} dp\hs\CH_*[p,n],}
where $\CH_*[p,n]$\ is the space of operators of momentum $p$\ at mass level
$n$.  These spaces are null unless $p^2=-2n$; and when this condition is
satisfied, ${\rm dim}\ \CH_b[p,n]={\rm dim}\ \CH_f[p,n]=p_8(n)$, since the
spectrum of the string is supersymmetric.

Before we write down any examples of physical operators, let us recall that the
spectrum of the $N=1$\ string admits an infinite number of inequivalent,
irreducible representations
\ref\FMS{D.  Friedan, E.  Martinec and S.  Shenker, ``Conformal Invariance,
Supersymmetry and String Theory," Nucl.  Phys.  {\bf B271}(1986)93.}.
These representations are labeled by their charge with respect to the field
$\phi$, where $\phi$\ enters into the bosonization of the $(\be,\ga)$\ system.
An operator with charge $n$\ is said to be in the $n$-th ghost picture, or to
carry ghost charge $n$.
Bosonic operators always carry integral ghost charge, and fermionic operators
half-odd integral charge.
We may pass between the different representations using the picture changing
operation:  $\CO_{q+1}=[Q_{BRST},2\xi\CO_q]$.
Since we will be only be interested in computing scattering amplitudes, it is
only necessary to keep the part of $Q_{BRST}$\ that conserves the $(b,c)$\
ghost charge, namely $Q_{BRST}=\oint dz\hs{1\over2}\psi\cdot\partial X
e^\phi\eta(z)$.
The $(-1)$- and $(-1/2)$-pictures are called the canonical pictures.
Operators take on their simplest form when written in one of these pictures.

The massless spectrum consists of a 10-dimensional target space vector and a
10-dimensional target space spinor.
These operators have representatives \FMS\ in the canonical pictures of the
form
\eqna\canon
$$\eqalignno{
V_{-1}(\ze,p)&=c\ze\cdot\psi_{(-1)}e^{ip\cdot X}=c\ze\cdot\psi
e^{-\phi}e^{ip\cdot X},&\canon a\cr
V_{-1/2}(u,p)&=c\bar u S_{(-1/2)}e^{ip\cdot X}=c\bar u S e^{-\phi/2}
e^{ip\cdot X}.&\canon b\cr\cr}$$
BRST invariance requires that $p^2=0$\ for both operators, $\ze\cdot p=0$ and
$\bar u p\cdot\ga=0$; and the GSO projection requires that $u$\ be a chiral
spinor.
We will also need the form of the bosonic vertex in the 0-ghost picture.
This is given by
\eqn\veczero{
V_0(\ze,p)=-c\ze\cdot(\partial X+ip\cdot\psi\psi)e^{ip\cdot X}.}
The general form of the physical operators at the first massive level has been
worked out by Koh {\it et al.} in
\ref\massive{I.  G.  Koh, W.  Troost and A.  Van Proyen, ``Covariant Higher
Spin Vertex Operators in the Ramond Sector," Nucl.  Phys.
{\bf B292}(1987)201}.
We will not list their results here, but rather simply make use of them as the
need arises.

\subsec{The Bracket}

We now want to extend the bracket to the $N=1$\ string.
Here the existence of an infinite number of inequivalent ghost pictures
immediately raises the following question:  does the bracket depend on which
picture we use?
Consider the derivation of the bracket relations, in particular the correlator
$\bigl<0\big|V_1(z_1)V_2(z_2)\ldots V_n(z_n)
\oint J\big|0\bigr>$.
It is always possible to redistribute the ghost charges inside a correlation
function, so let us imagine shifting one unit of ghost charge from some $V_i$\
to a different $V_j$, both before and after we perform the contour deformation.
Since the resulting relations should be the same, this suggests that taking the
bracket of an operator $\CO_1$\ with the picture changed version of another
operator $\CO_2$ should be the same as the picture changed version of
$\{\CO_1,\CO_2\}$.
For a more direct proof of this, consider the expression $[Q_{BRST},
2\xi\{\CO_1,\CO_2\}]$.
Since neither $\CO_1$\ nor $\CO_2$\ contains $\eta$, we can move $2\xi$\ inside
the bracket next to $\CO_2$.
Furthermore since $\CO_1$\ is a always assumed to be a physical operator,
$Q_{BRST}$\ sees only $2\xi\CO_2$.
Thus we can move $Q_{BRST}$\ inside the bracket, leaving us with the desired
expression $\{\CO_1,[Q_{BRST},2\xi\CO_2]\}$.
Using the relation $\{\CO_1,\CO_2\}=-(-1)^{q\cdot p}\{\CO_2,\CO_1\}$, we can
also write this as $\{[Q_{BRST},2\xi\CO_1],\CO_2\}$.
Thus the picture changing operation commutes with the bracket.
Along these same lines, the argument that says we can redistribute the ghost
charges inside a correlation function \FMS\
also tells us that we can redistribute the ghost charge inside the bracket.
These two properties tell us that the bracket is independent of the ghost
pictures of the operators involved.
We will see below that this is indeed borne out by example.

Having shown that the bracket is well-defined, we want to compute the brackets
needed to write down the relations among scattering amplitudes involving only
massless operators.
Because the bracket is picture independent, we will choose the pictures such
that the brackets take on a simple form.
The bracket of two massless bosonic operators, with one in the
$(-1)$-picture and the other in the 0-picture, whose momenta $q$\ and $p$\
satisfy $q\cdot p=0$, is given by
$$\eqalign{\{V_{-1}(\ze,q),V_0(\ze',p)\}
&=-\oint dw\hs\ze\cdot\psi_{(-1)}e^{iq\cdot X(w)}
c\ze'\cdot(\partial X+ip\cdot\psi\psi)e^{ip\cdot(z)}\cr
&=ic(\ze\cdot p\ze'-\ze'\cdot q\ze-\ze\cdot\ze' p)\cdot\psi_{(-1)}
e^{i(q+p)\cdot X(z)}\cr
&=i\ze\cdot p V_{-1}(\ze',q+p)-i\ze'\cdot q V_{-1}(\ze,q+p)\cr
&\qquad\qquad -i\ze\cdot\ze' V_{-1}(p,q+p).\cr}$$
To verify that this result is indeed picture independent, let us compute the
bracket with both operators in the $(-1)$-picture.
$$\eqalign{\{V_{-1}(\ze,q),V_{-1}(\ze',p)\}(z)
&=\oint dw\hs\ze\cdot\psi_{(-1)}e^{iq\cdot X(w)}
c\ze\cdot\psi_{(-1)}e^{iq\cdot X(z)}\cr
&=c(\ze\cdot\psi\ze'\cdot\psi-i\ze\cdot\ze' q\cdot\partial X)
e^{-2\phi}e^{i(q+p)\cdot X(z)}.\cr}$$
It is easy to check that the picture changed version of this result is
the same as above.
Because of this independence we will write
\eqn\brbos{
\eqalign{\{V_b(\ze,q),V_b(\ze',p)\}
&=i\ze\cdot p V_b(\ze',q+p)-i\ze'\cdot q V_b(\ze,q+p)\cr
&\qquad\qquad -i\ze\cdot\ze' V_b(p,q+p),\cr}}
with the understanding that the ghost picture will be chosen so that the ghost
charges balance.

The mixed case of the bracket between a massless boson and a massless fermion
of momenta $q$\ and $p$, again with $q\cdot p=0$, is
\eqn\brmix{
\{V_b(\ze,q),V_f(u,p)\}
=-{i\over2}V_f\bigl((\qsl+\psl)\zesl u,q+p\bigr).}
It is easy to check that both $\{V_0(\ze,q),V_{-1/2}(u,p)\}$\ and
$\{V_{-1}(\ze,q),V_{1/2}(u,p)\}$, and the picture changed version of
$\{V_{-1}(\ze,q),V_{-1/2}(u,p)\}$\ all give this answer.
Finally the bracket between two massless fermionic operators, again with
$q\cdot p=0$, is given by
\eqn\brferm{
\{V_f(v,q),V_f(u,p)\}=-{1\over\sqrt2}V_b(\bar v\ga u,q+p).}

\subsec{The Massless Relations}

Now let us use these brackets to derive relations among scattering amplitudes
involving only massless operators.
We begin with the relation generated by $n+1$\ massless bosonic operators
$\{V_b(\ze_1,p_1),\allowbreak \ldots,V_b(\ze_n,p_n);V_b(\ze,q)\}$, whose
momenta satisfy $q\cdot p_i=0$\ for all $p_i$.
Again since the bracket is picture independent, we will always specify the
operators by giving the spinors, polarization tensors, {\it etc.} needed to
characterize them in the canonical picture, with the understanding that the
ghost charges will be chosen so that they sum to $-2$.
The relation associated to these operators is
\eqn\brelbos{
\eqalign{
&\sum_{i=1}^n\ze\cdot p_i
\CA_{b\cdots b}\left(\matrix{\cdots&\ze_i&\cdots\cr
\cdots&q+p_i&\cdots\cr}\Big|s_{ij}\right)+
\sum_{i=1}^n\ze_i\cdot q
\CA_{b\cdots b}\left(\matrix{\cdots&\ze&\cdots\cr
\cdots&q+p_i&\cdots\cr}\Big|s_{ij}\right)\cr
&\hskip 20pt+\sum_{i=1}^n\ze\cdot\ze_i
\CA_{b\cdots b}\left(\matrix{\cdots&p_i&\cdots\cr
\cdots&q+p_i&\cdots\cr}\Big|s_{ij}\right)=0.\cr}}

\noindent
The finite difference relation among scattering amplitudes involving only
massless fermions can be derived from $\{V_f(u_1,p_1),\ldots,
V_f(u_n,p_n);V_b(\ze,q)\}$, where again $q\cdot p_i=0$.
\eqn\brelferm{
\sum_{i=1}^n \CA_{f\cdots f}\left(\matrix
{\cdots&(\qsl+\psl_i)\zesl u_i&\cdots\cr
\cdots&q+p_i&\cdots\cr}\Big|s_{ij}\right)=0.}

\noindent
Finally we can find relations among scattering amplitudes involving both
massless bosons and fermions by choosing the generating operator to be
fermionic.
For example the relation associated with $\{V_f(u_1,p_1),
V_f(u_2,p_2),V_f(u_3,p_3),\allowbreak V_b(\ze_4,p_4);V_f(v,q)\}$\ is
\eqn\brelmix{
\eqalign{
\noalign{\vskip 3pt}
&{\cal A}_{ffff}\left(\matrix
{u_1&u_2&u_3&(\qsl+\psl_4)\zesl_4 v\cr
p_1&p_2&p_3&q+p_4\cr}\Big|s,t\right)\cr
\noalign{\vskip 3pt}
&\hskip 5pt +{\cal A}_{ffbb}\left(\matrix
{u_1&u_2&\bar v\ga u_3&\ze_4\cr
p_1&p_2&q+p_3&p_4\cr}\Big|s,t\right)
+{\cal A}_{fbfb}\left(\matrix
{u_1&\bar v\ga u_2&u_3&\ze_4\cr
p_1&q+p_2&p_3&p_4\cr}\Big|s,t\right)\cr
\noalign{\vskip 3pt}
&\hskip 5pt +{\cal A}_{bffb}\left(\matrix
{\bar v\ga u_1&u_2&u_3&\ze_4\cr
q+p_1&p_2&p_3&p_4\cr}\Big|s,t\right)=0.\cr}}

\noindent
Since $V_f$\ is the supersymmetry operator at non-vanishing momentum, this
should be thought of as a generalized supersymmetry relation.
For general $n$, there are $[(n+1)/2]$\ such relations.
Since the four possible 4-point scattering amplitudes have all been computed
\GSW, it is possible to check these relations for $n=4$, and indeed they hold.
This provides a non-trivial check on our formalism.

\subsec{The Massive Relations}

Having shown that Moore's ideas can be extended to the $N=1$\ string, we want
to prove that one of his non-trivial assertions, namely that we can express any
$n$-point amplitude in terms of the $n$-point tachyon amplitude, has a
counterpart in the superstring.
Consider the set of $n$ operators $V_b^l=V_b(\ze,lq)$, $l=1,\ldots,n$, where
$\ze\cdot q=q^2=0$.
Combining the bracket with these operators, we have the maps
\eqn\ddf{
\{V_b^l,*\}:{\cal H}_*[p-(n-l)q,n-l]\rightarrow
{\cal H}_*[p-nq,n],}
where we have assumed $q\cdot p=0$.
If we then note that these maps are equivalent to the action of the bosonic DDF
operators $\ze\cdot A_l$, then the no-ghost theorem implies that the map had by
summing over $l$\ is onto \GSW.
This means that given, say, any $m$-point bosonic scattering amplitude
involving string states at levels $n_1,\ldots,n_m$, we can always find an $l$
such that the relation generated by $\{V_b(n_1-l),\ldots,
V_b(n_m);V_b^l\}$\ expresses our original amplitude in terms of amplitudes
involving at least one string states at a strictly lower mass level.
Repeating this process we can express any bosonic $m$-point function in terms
of the $m$-point amplitude for the massless bosonic states.
This argument easily generalizes to amplitudes involving an even number of
fermionic states.
Thus we can say that an arbitrary amplitude can be expressed in terms of one of
a finite number of massless amplitudes.
For the first few cases $n=4,5,6,7$\ there are na\"\i vely $4,4,8,10$\
amplitudes.
Using the generalized supersymmetry relations, we can reduce this to $2,2,5,7$\
independent amplitudes.

\newsec{Recursion Relations}

\subsec{An Initial Attempt}

We would next like to derive a set of recursion relations for the massless
scattering amplitudes that would allow us to evaluate them without ever
having to go through the usual conformal field theory calculations.
We will restrict ourselves to the simplest case of 4-point amplitudes.
Following the pattern of Moore's original work, we are tempted to proceed as
follows.
We choose the generating current to be $V_b(\ze,q)$.
This guarantees that we will not change the number of either bosons or
fermions.
We then choose the momenta such that one of the amplitudes involves a massive
string state.
To see why this is necessary, note that in the massless relations derived
above, all of the amplitudes are evaluated at the same values of $s$\ and $t$.
These sort of relations are useless if we want to have a recursive method for
finding the dependence of the scattering amplitudes on the kinematical
invariants.
Choosing the momenta such that $q\cdot p_i\ne 0$\ for some $p_i$\ guarantees
that we will have different values of $s$\ and $t$.
Finally we try to choose the polarization vectors and spinors such that the
massive state is BRST trival, leaving us again with a relation among only
massless amplitudes.

We will begin with $\{V_f(u_1,p_1),\ldots,V_f(u_4,p_4); V_b(\ze,q)\}$, with the
momenta chosen such that $p_1\cdot q=1$, $p_2\cdot q=p_3\cdot q=0$\ and
$p_4\cdot q=-1$.
This requires computing $\{V_{-1}(\ze,q),V_{-1/2}(u,p)\}$
for $q\cdot p=-1$.
This is given by
\eqn\masferm{
{1\over\sqrt2}(Y\cdot\partial X+R\cdot\psi\psisl)
S_{(-3/2)}e^{i(q+p)\cdot X(z)},}
where we have defined $Y=iq\bar u\zesl$\ and $R={1\over4}\bar
u\ze-{1\over72}\bar u\zesl\ga$.
Note that this is the vertex operator for a {\it massive} string state.
Now we want to choose $\ze$\ and possibly a number of the $u_i$\ such that that
massive vertex is BRST trivial.
One of the two possible sets of conditions that the render the vertex trivial
is given by the pair of equations \massive
\eqn\brstcond{
\eqalign{8R+Y(\qsl+\psl)+\Rsl\ga&=0,\cr
Y-\Rsl(q+p)-4R(\qsl+\psl)-{1\over9}\Ysl\ga&=0.\cr}}
For our particular $Y$\ and $R$, the first of these simplifies to $2\bar
u\ze+q\bar u
\zesl(\qsl+\psl)=0$.
The only solution of this equation is $\ze=q$,
and in this case all of the other states generated by $V_b(\ze,q)$\ vanish
identically.
Thus we fail to generate a nontrivial relation.
There is another set of conditions on $Y$\ and $R$\ that also yields BRST
trivial states, but these require that $Y\propto q+p$, so we can never write
our state as one of these.

Let us try again with $\{V_b(\ze_1,p_1),\ldots,
V_b(\ze_4,p_4);V_b(\ze,q)\}$,
$p_1\cdot q=1$, $p_2\cdot q=p_3\cdot q=0$\ and $p_4\cdot q=-1$.
We begin by computing $\{V_{-1}(\ze,q),V_{-1}(\ze',p)\}$\ for
$q\cdot p=-1$:
$$\left\{\ze\cdot\psi\ze'\cdot\psi iq\partial X-
{1\over2}\ze\cdot\ze'\left(iq\cdot\partial^2 X+(iq\cdot\partial X)^2\right)
+\ze\cdot\partial\psi\ze'\cdot\psi\right\}
e^{-2\phi}e^{i(q+p)\cdot X}$$
If we apply the picture changing operator to this we find
\eqn\masbos{
\eqalign{
&\bigl(\al_{\mu\nu\rho}\psi^\mu\psi^\nu\psi^\rho+
(\si_{\mu\nu}+\al_{\mu\nu})\partial X^\mu\psi^\nu+
\si_\mu\partial\psi^\mu\bigr)e^{-\phi}e^{i(q+p)\cdot X}\cr}}
where
\eqn\mascoef{
\eqalign{
\al_{\mu\nu\rho}&=-{i\over 6}q_{[\mu}\ze_\nu\ze'_{\rho]}\cr
\si_{\mu\nu}+\al_{\mu\nu}&=-\ze\cdot p\ze'_\mu q_\nu+
\ze\cdot q\ze'_\mu q_\nu-\ze\cdot\ze' q_\mu q_\nu-\ze'_\mu\ze_\nu\cr
\si_\mu&=i\ze\cdot\ze' q_\mu-i\ze'\cdot q\ze_\mu\cr}}
Here $\si_{\mu\nu}$\ denotes the symmetric piece and $\al_{\mu\nu}$\ the
antisymmetric.
If we choose $\ze=q$, then this vertex operator vanishes.
However, as in the case of four fermions, all of the verticies generated from
$V_b(\ze,q)$\ in this case will be BRST trivial.
Since we cannot set $\ze=q$, we must have a nontrivial $\si_\mu$.
{}From \massive\ we know that $\si_\mu$\ contributes only to the BRST trivial
part of a vertex.
Thus we should separate out the contribution from $\si_\mu$\ and check whether
there is any remaining physical piece that cannot be removed.
First of all $\si_\mu$\ must satisfy $\si\cdot(p+q)=0$.
This condition can be written as $\ze\cdot\ze'+\ze'\cdot q\ze\cdot p=0$.
The BRST trivial piece of $\si_{\mu\nu}$\ is $\si_{\mu\nu}^{trivial}=
(q+p)_{(\mu}\si_{\nu)}$.
Subtracting this from the above expression for $\si_{\mu\nu}$, we find
\eqn\masphys{
\eqalign{
\si_{\mu\nu}^{phys}&=
-{1\over2}\ze\cdot p(\ze'_\mu q_\nu+q_\mu\ze'_\nu)-
{1\over2}(\ze'_\mu\ze_\nu+\ze_\mu\ze'_\nu)\cr
&\hskip 20pt +{1\over2}\ze\cdot\ze'(p_\mu q_\nu+q_\mu p_\nu)-
{1\over2}\ze'\cdot q(p_\mu\ze_\nu+\ze_\mu p_\nu).\cr}}
With the help of the the above condition we can show that
$\si_{\mu\nu}^{phys}$\ satisfies $(q~+~p)^\mu\si_{\mu\nu}^{phys}=0$\ and
$\eta^{\mu\nu}\si^{phys}_{\mu\nu}=0$, so that $\si_{\mu\nu}^{phys}$ is indeed a
phyical state \massive.
The upshot of this is that we cannot arrange things such that the massive
vertex is BRST trivial, and we again fail to find a recursion relation.

\subsec{A Generalized Bracket}
{}From what we have seen above, it would appear impossible to determine any of
the 4-point amplitudes using only the bracket relations.
What we want to show now is that we have been too narrow in our thinking.
Since the bracket was originally conceived as a special case of the
Gerstenhaber bracket, we assumed that it could only be applied to the physical
operators of the $N=1$\ string.
However there is no reason we cannot consider
$\{\CO_1,\CO_2\}$, where $\CO_2$\ is a physical operator but $\CO_1$\ is just a
dimension one chiral operator constructed from the fields of our theory, as
long as the result is again a physical operator.
${\cal O}_1$\ must still have dimension one, since otherwise our contour
deformation arguments would not go through.
To show that there are indeed examples of such operators, let us compute the
bracket of a massless fermion and what can be thought of as an on-shell,
bosonic string tachyon.
\eqn\brtacferm{
\eqalign{
\{V_T(q),V_{-1/2}(u,p)\}&=\oint dw\hs e^{iq\cdot X(w)}c\bar u S_{(-1/2)}
e^{ip\cdot X(z)}\cr
&=V_{-1/2}(u,q+p),\cr}}
where we have assumed that their momenta satisfy $q\cdot p=-1$.
If we replace the fermion by a boson, we find the analagous result
\eqn\brtacbos{
\{V_T(q),V_{-1}(\ze,p)\}=V_{-1}(\ze,q+p),}
where again $q\cdot p=-1$.

Since the effect of the tachyon is to shift the momentum of the fermion, this
would appear to be just what we need to generate recursion relations.
Thus let us consider the relation associated with
$\{V_{-1/2}(u_1,p_1),V_{-1/2}(u_2,p_2),\allowbreak
V_{-1/2}(u_3,p_3),V_{-1/2}(u_4,p_4);V_T(q)\}$, where
$q\cdot p_1=q\cdot p_2=q\cdot p_3=-q\cdot p_4=1$.  We find
\eqn\trirel{
\CA_{ffff}(s-1,t-1)=\CA_{ffff}(s-1,t)+\CA_{ffff}(s,t-1).}
This is just the kind of relation found by Moore for the 4-point tachyon
amplitude for the bosonic string \bracket.
As before it is easy to show that this relation is satisfied.
Note that in all of the above computations, we never used the picture
independent notation of section 3, but rather made it clear that we were always
working in the canonical ghost pictures.
This was done for two reasons.
First of all our arguments for picture independence do not go through if both
operators in the bracket are not physical.
Thus we had to restrict ourselves to some definite picture.
Second of all, had we worked in say the (+1/2)- and 0-pictures, we would have
found that the tachyon did {\it not} map physical states into physical states.
The tachyon can be used in the canonical pictures because there the massless
states are the product of a tachyon and something that the tachyon does not
see.

As with the tachyon amplitude, equation \trirel\ is not enough to solve for the
4-point fermionic amplitude.
Thus let us continue along the path followed by Moore, and consider the bracket
of a massless fermionic operator and what would be a physical photon of the
bosonic string.
\eqn\brphferm{
\eqalign{
\{V_\ga(\ze,q),V_{-1/2}(u,p)\}&=\oint dw\hs\ze\cdot\partial X
e^{iq\cdot X(w)}\bar u S_{-1/2}e^{ip\cdot X(z)}\cr
&=\left\{\vcenter{\openup1\jot
\halign{$\hfil#$&${}#\hfil$&\hskip10pt$\hfil#$&${}#\hfil$\cr
&-i\bigl(\ze+(\ze\cdot p)q\bigr)\cdot p V_{-1/2}(u,q+p)&q\cdot p&=0\cr
&\bigl(\ze+(\ze\cdot p)q\bigr)\cdot\partial X \bar u S_{(-1/2)}
e^{i(q+p)\cdot X}&q\cdot p&=-1\cr}}
\right.\cr}}
Since $q^2=0$, we can define $\ze'=\ze+(\ze\cdot p)q$\ and still preserve the
relation $\ze'\cdot q=0$.
This greatly simplifies the bracket relations that follow.
Now consider the relation generated by
$\{V_{-1/2}(u_1,p_1),\allowbreak V_{-1/2}(u_2,p_2),\allowbreak
V_{-1/2}(u_3,p_3),\allowbreak V_{-1/2}(u_4,p_4);V_\ga(q)\}$, where
$q\cdot p_1=q\cdot p_2=0$ and $-q\cdot p_3=q\cdot p_4=1$.
\eqn\brelmas{
\CA_{ff\ze f}(s,t)=-i\ze\cdot p_1\CA_{ffff}(s,t)
-i\ze\cdot p_2\CA_{ffff}(s,t-1),}
where the subscript $\ze$\ on the left hand side of the relation denotes the
massive state $\ze\cdot\partial X\bar uS_{(-1/2)}e^{i(q+p)\cdot X(z)}$.
If by some convenient choice of $\ze$\ we could render this state BRST trivial,
equation \brelmas\ would reduce to the recursion relation necessary to
determine $\CA_{ffff}$.
But according to Koh {\it et al.} \massive\ there is no value of $\ze$\ for
which this state is BRST trivial.
Thus we seem to be in the same straits as before.
However since the form of this operator is so much simpler than what we
previously encountered for massive states, let us blindly proceed for the
moment.

We begin by imposing the BRST invariance conditions for the massive state.
These imply that $\ze\cdot(p_3+q)=\bar u_3\qsl=\bar u_3\zesl=0$.
Recall that we also have the initial conditions $q\cdot p_3=-1$,
$q^2=p_3^2=\ze\cdot q=\bar u_3\hskip 1pt\psl_3=0$.
Now $\CA_{ffff}(s,t)$\ must be of the form
\eqn\genform{
f(s,t)\bar u_1\ga u_2\cdot\bar u_3\ga u_4
+g(s,t)\bar u_2\ga u_3\cdot\bar u_4\ga u_1}
since $\ga_{\al\be}\cdot\ga_{\de\ep}$\ and $\ga_{\be\de}\cdot\ga_{\ep\al}$\ are
the only two independent $SO(10)$\ invariants with four spinor indicies.
Let us set $u_2=u_3$.
This restricts our attention to the unknown function $f(s,t)$, and also implies
that $\bar u_3\hskip 1pt\psl_2=0$.
Using $\bar u_3\zesl=0$\ along with the other conditions on $u_3$, we find that
the most general form of $\ze$\ is $\ze=q+\al p_2+\be p_3$.
Using $\ze\cdot(q+p_3)=0$\ we can then solve for $\be$\ to find $\ze=q
+\al p_2+(\al t-1)p_3$.
Thus we are left with a one-parameter family of massive states.
Now let us recall that we want to get rid of the massive state.
Since we cannot choose $\ze$\ such that the massive state is BRST trivial,
perhaps we can choose $\al$\ such that the amplitude involving the massive
state vanishes.

The piece of the scattering amplitude involving only $X$\ and its derivatives
is given by
$$\eqalign{
&\bigl<e^{ip_1\cdot X(z_1)}e^{ip_2\cdot X(z_2)}\ze\cdot\partial X
e^{ip_3\cdot X(z_3)}e^{ip_4\cdot X(z_4)}\bigr>\cr
&\qquad=\left({{\ze\cdot p_1}\over z_3-z_1}+{{\ze\cdot p_2}
\over z_3-z_2}+{{\ze\cdot p_4}\over z_3-z_4}\right)
\bigl<e^{ip_1\cdot X(z_1)}e^{ip_2\cdot X(z_2)}
e^{ip_3\cdot X(z_3)}e^{ip_4\cdot X(z_4)}\bigr>\cr}$$
If we fix the $SL(2,\IC)$\ symmetry as usual, then this reduces to
$$\left(-{{\ze\cdot p_2}\over 1-x}+{{\ze\cdot p_4}\over x}\right)
x^s(1-x)^t,$$
where we have also included the contributions of the $(b,c)$\ ghost system.
If we note that the remaining contribution to the scattering amplitude does not
contain any explicit factors of $s$\ or $t$, then we can think of the
denominators $x$\ and $1-x$\ as effectively shifting $s$\ and $t$; and
requiring this amplitude to vanish implies that $f(s,t)$\ must satisfy the
additional relation
\eqn\masvan{\ze\cdot p_2 f(s,t-1)=\ze\cdot p_4 f(s-1,t)}
Since $\ze$\ depends on only a single parameter, with this equation we should
have enough relations both to find the needed $\al$ and to solve for $f(s,t)$.

Assuming that the desired $\al$\ does indeed exist, equation \brelmas\ becomes
\eqn\rrintone{
\ze\cdot p_1 f(s,t)=-\ze\cdot p_2 f(s,t-1).}
Combining this with equation \masvan\ we find
\eqn\rrinttwo{
\ze\cdot p_1 f(s,t)=-\ze\cdot p_4 f(s-1,t).}
If we use these equations to write the right hand side of \trirel\ in terms of
$f(s-1,t-1)$, the resulting equation allows us to solve for $\al$, with the
result $\al=1/(s+t)$.
With this value of $\al$\ equations \rrintone\ and \rrinttwo\ become
\eqn\recrel{
f(s,t)={t\over s+t}f(s,t-1)={{s-1}\over s+t}f(s-1,t).}
Using these recursion relations we find
\eqn\fsol{
f(s,t)=c_4{{\Gamma(s)\Gamma(t+1)}\over\Gamma(s+t+1)}}
in agreement with the known result.
Of course our relations only allow us to determine $f(s,t)$\ for integer $s$\
and $t$, but as in \bracket\ we can analytically continue our result to the
entire complex plane by using two mild assumptions about the analyticity of the
amplitudes.

In order to solve for the unknown function $g(s,t)$, we simply repeat the
entire process, setting $u_1=u_2$\ rather than $u_2=u_3$.
This yields $g(s,t)=c'_4\Gamma(s)\Gamma(t+1)/\Gamma(s+t+1)$, again in agreement
with the known result.
Finally we can use the cyclical symmetry of the amplitude to fix $c'_4=c_4$.
Our generalized bracket can also be used to compte $\CA_{bfbf}$, and then the
two relations of the form \brelmix\ can be used to determine the two remaining
amplitudes $\CA_{bbff}$\ and $\CA_{bbbb}$.
This completes the solution of the 4-point massless amplitudes using only the
bracket.

\subsec{Extension to $n>4$}

Let us begin with the $2n$-point massless fermionic amplitude for $n>3$.
The generalizations of equation $\trirel$\ are generated by
$\{V_{+1/2}(p_1),\ldots,V_{+1/2}(p_{n-3}),\allowbreak
V_{-1/2}(p_{n-2}),\ldots,V_{-1/2}(p_{2n});
V_T(q)\}$, with $q\cdot p_i=1$\ for $i=1,\ldots,n-1$\ and $q\cdot p_i=-1$\ for
$i=n,\ldots,2n$, and its various permutations.
Note that we omit the polarization tensors for simplicity's sake.
These relations should serve as the analogues of Moore's triangle relations
\bracket, which serve to reduce the $2n$-point amplitude from a function of
$n(2n-3)$\ variables to one of $2n-3$\ variables.
The generalizations of \brelmas\ are the relations generated by
$\{V_{+1/2}(p_1),\ldots,V_{+1/2}(p_{n-3}),
V_{-1/2}(p_{n-2}),\ldots,V_{-1/2}(p_{2n});
V_\gamma(q)\}$\ with $q\cdot p_i=1$\ for $i=1,\ldots,n$\ and $q\cdot p_i=-1$\
for $i=n+1,\ldots,2n$, and its permutations, along with the relations
constraining the amplitudes involving massive states to vanish.
These would then be used to solve for the dependence on the remaining
variables.
The six point amplitude has in fact been worked out in
\ref\sixpoint{V.  A.  Kostelecky, O.  Lechtenfeld, S.  Samuel, K. Verstegen, S.
Watamura, D. Sahdev, ``The Six-Fermion Amplitude in the Superstring," Phys.
Lett.  {\it 183B}(1987)299.} using a rather involved technique developed in
\ref\diagmeth{V.  A.  Kostelecky, O.  Lechtenfeld, W. Lerche, S.  Samuel and S.
Watamura, ``Conformal Techniques, Bosonization and Tree-Level String
Amplitudes," Nucl. Phys {\bf B288}(1987)173.}
for computing correlators.
It would be an interesting exercise to try to use the above bracket relations
to reproduce their results

Similar relations to these can also be generated for amplitudes involving
either one or two bosonic operators.
However, when we try to use the tachyon to generate relations for amplitudes
with three or more bosons, we find that it is impossible to do so.
Now as we saw in the previous subsection, one of the problems with using the
generalized bracket is that it is not picture independent.
For the particular operators we chose to use, $V_T$\ and $V_\ga$, we had to
restrict ourselves to the canonical picture.
When we couple this restriction with the necessity of having to choose
representatives of operators whose ghost charges sum to $-2$, we find that the
two requirements are sometimes incompatible.
For example in the case of three or more bosons, no matter how we choose the
ghost pictures and the momenta, we end up having to compute the bracket of the
tachyon with a state in a picture other than the canonical picture, which as we
mentioned does not give us back a physical state.
When we have four or more bosons, we can no longer even use the photon.
One possible way of avoiding these restrictions would be to use higher mass
states from the bosonic string to generate relations.
Another possibility would be to allow the tachyon and photon to generate higher
mass states of the $N=1$\ string.
While the former approach seems quite tenable, the latter is hampered by the
fact that the BRST analysis for the $N=1$\ string has not yet been perfomed at
even the second massive level.

\newsec{Lifting the Restriction $n\le10$}

In his original work \bracket\ Moore found that he could only apply his
bracket relations to scattering amplitudes involving at most tweny-six string
states.
The origin of this restriction was explained in section 2.
In \addendum\ he showed how this restriction could effectively be lifted.
This was done as follows.
The general form of the restriction on $n$\ is that it be no greater than the
number of uncompactified target space dimensions.
Thus the idea that immediately presents itself is to embed our theory in
another with a larger target space.
The problem with this is that doing so takes us off criticality.
However we can restore the central charge to zero by tensoring our enlarged
theory with a number of ghost systems.

To be more exact, let us denote by $C(M)$\ the conformal field theory with
target space $M$, {\it e.g.} $C(\IR^{1,25})$\ is the theory that underlies the
ordinary critical bosonic string.
Moore embedded the theory based on $C(\IR^{1,25})$\ into that based on
$C(R^{1+E,25+E})\bigotimes_{i=1}^E[<\xi_i,\eta_i>\cap\hskip 2pt
{\rm ker}(\oint\eta_i)]$, where $<\xi_i,\eta_i>$\ is a spin $(0,1)$\ fermionic
ghost system.
This allowed him to compute amplitudes involving up to $26+2E$\ string states,
and since $E$\ is arbitrary, this effectively removed the restriction on $n$.
Here we want to show that the same procedure also works for the $N=1$\ string
if we extend the ghost fields to $N=1$\ superfields.
We denote by $SC(M)$\ the superconformal field theory (SCFT) with target space
$M$.
Since the $N=1$\ string is a SCFT with target space $\IR^{1,9}$, in order to
apply Moore's construction we need to extend not only the bosonic target space
fields but also their fermionic partners.
To cancel the contribution to the central charge from these extra fields, we
will tensor our theory with an additional $E$ copies of a spin $(1/2,1/2)$\
bosonic ghost system $<\tilde\xi,\tilde\eta>$.
Note that this system does not have a $U(1)$\ anomaly, so there is no need to
restrict ourselves to a subspace of this theory.
We can combine these two ghost systems into a single superghost system
described by $\Xi=\xi+\theta\tilde\xi$\ and $H=\tilde\eta+\theta\eta$,
and thus our full theory may be written as
$SC(\IR^{1+E,9+E})\bigotimes_{i=1}^E[<\Xi_i,H_i>\cap\hskip 2pt
{\rm ker}(\oint\int d\theta H_i)]$.

Once we have solved for the amplitudes in this extended theory, how do we pull
out the amplitudes for our original theory?
We begin by restricting ourselves to amplitudes involving states built only
from the ordinary (supersymmetric) matter fields.
Since the bracket is closed with respect to such states, we need not worry
about states involving ghost fields propagating in any of the intermediate
channels; after all the coeffecients of the bracket relations are nothing more
than the three-point functions.
Having computed these amplitudes, which live in a $10+2E$\ dimensional target
space, we simply continue our results to momenta whose last $2E$\ components
vanish.

\newsec{Discussion}

In this paper we have succeeded in extending most of Moore's original analysis
to the the $N=1$\ string, thus doing the second of his things that ``we should
do."
This required a slight extension of the notion of the bracket to include the
bracket between a physical operator and an ordinary dimension one chiral field.
Another way of thinking about this extension is to say that the $N=1$\ string
admits an algebra of external operators, or that the states of the $N=1$\
string (in the canonical picture) form a module over this external algebra.
The most glaring shortcoming of the present work is a lack of a general proof
that the massless amplitudes are computable entirely in terms of the bracket.
However, being that the 4-point amplitude is computable, we have every
confidence that the general amplitude should be so as well.

Having shown that the bracket can be extended to the $N=1$\ string, we can now
begin to ask questions about more interesting theories such as the type II
superstring and the heterotic string.
Furthermore the bracket should be applicable not only to the flat backgrounds
studied up until now, but also to more general conformal field theories as long
as they contain some noncompact sector.
In particular we could consider the above mentioned theories, but compactified
on some non-trivial, internal conformal field theory; or we could even consider
non-critical theories.
In any of these examples it would be interesting to see to what extent the
bracket determines the S-matrix.

\bigskip
\centerline{\bf Acknowledgements}
I would like to thank M.  Bershadsky for suggesting that I consider the
extension of Moore's work to the $N=1$\ string and for numerous discussions
throughout the progress of this work.
This work is supported by an NDSEG Graduate Fellowship and in part by the
Packard Foundation and the NSF per grant PHY-92-18167.

\listrefs

\end